\documentclass[12pt]{article}

\usepackage{amsfonts}
\usepackage{amsmath}
\usepackage{epsfig}
\usepackage{latexsym}
\usepackage{graphicx}
\usepackage{epstopdf}
\usepackage{color}
\usepackage{amssymb}
\usepackage{framed}

\numberwithin{equation}{section}
\allowdisplaybreaks

\def\spa#1{\phantom{\fbox{\rule[-#1cm]{0cm}{0cm}}}}

\def\be{\begin{equation}}
\def\ee{\end{equation}}
\def\bea{\begin{eqnarray}}
\def\eea{\end{eqnarray}}
\def\bequ{\begin{equation}}
\def\eequ{\end{equation}}

\def\del{\partial}

\renewcommand{\thefootnote}{\fnsymbol{footnote}}

\newcommand{\eq} {equation}
\newcommand{\eqa} {eqnarray}
\newcommand{\NN} {\nonumber}

\begin{document}
\hfuzz=100pt
\title{Quantum Black Hole Entropy\\
from 4d Supersymmetric Cardy formula}
\author{
 Masazumi Honda\footnote{mh974ATdamtp.cam.ac.uk} 
  \spa{0.5} \\
\\
{\small{\it Department of Applied Mathematics and Theoretical Physics,}}
\\ {\small{\it Centre for Mathematical Sciences, Wilberforce Road, Cambridge, CB3 0WA, UK}} \\
}
\date{\small{January 2019}}
\maketitle
\thispagestyle{empty}
\centerline{}

%%%%%%%%%%%%%%%%%%%%%%%%%%%%%%%%%%%%%%
\begin{abstract}
We study supersymmetric index of 4d $SU(N)$ $\mathcal{N}=4$ super Yang-Mills theory on $S^1 \times M_3$.
We compute asymptotic behavior of the index
in the limit of shrinking $S^1$ for arbitrary $N$
by a refinement of supersymmetric Cardy formula.
The asymptotic behavior for the superconformal index case ($M_3 =S^3$) at large $N$ agrees 
with the Bekenstein-Hawking entropy of rotating electrically charged BPS black hole in $AdS_5$
via a Legendre transformation as recently shown in literature.
We also find that the agreement formally persists for finite $N$ 
if we slightly modify the AdS/CFT dictionary between Newton constant and $N$.
This implies 
an existence of non-renormalization property of the quantum black hole entropy.
We also study the cases with other gauge groups and additional matters,
and the orbifold $\mathcal{N}=4$ super Yang-Mills theory.
It turns out that
the entropies of all the CFT examples in this paper are given by 
$2\pi \sqrt{Q_1 Q_2 +Q_1 Q_3 +Q_2 Q_3 -2c(J_1 +J_2 )} $
with charges $Q_{1,2,3}$, angular momenta $J_{1,2}$ and central charge $c$.
The results for other $M_3$ make predictions to the gravity side.
\end{abstract}
\vfill
\noindent
%DAMTP

\renewcommand{\thefootnote}{\arabic{footnote}}
\setcounter{footnote}{0}

\newpage
\setcounter{page}{1}
%\tableofcontents

%%%%%%%%%%%%%%%%%%%%%%%%%%%%%%%%%%%%%%%%%%%%%%%%%%
%%%%%%%%%%%%%%%%%%%%%%%%%%%%%%%%%%%%%%%%%%%%%%%%%%
%%%%%%%%%%%%%%%%%%%%%%%%%%%%%%%%%%%%%%%%%%%%%%%%%%
\section{Introduction}
%%%%%%%%%%%%%%%%%%%%%%%%%%%%%%%%%%%%%%%%%%%%%%%%%%
%%%%%%%%%%%%%%%%%%%%%%%%%%%%%%%%%%%%%%%%%%%%%%%%%%
%%%%%%%%%%%%%%%%%%%%%%%%%%%%%%%%%%%%%%%%%%%%%%%%%%
Since string theory is the candidate of consistent quantum gravity,
string theory should give microscopic explanation of black hole entropy \cite{Bekenstein:1972tm}.
%\cite{Bekenstein:1972tm,Bekenstein:1973ur,Bekenstein:1974ax,Hawking:1974sw,Hawking:1974rv}.
As well known, the seminal paper \cite{Strominger:1996sh} by Strominger and Vafa has derived
the Bekenstein-Hawking entropy of asymptotically flat black hole by counting BPS states in string theory.

In the context of AdS/CFT \cite{Maldacena:1997re},
%\cite{Maldacena:1997re,Gubser:1998bc,Witten:1998qj},
this problem is mapped into 
whether an entropy of an asymptotically AdS black hole is explained by counting states of a dual CFT.
Recently there has been great steps to understand this problem along two directions.
First,
the black hole entropies of static dyonic BPS black holes
has been reproduced by topologically twisted indices of 3d $\mathcal{N}=6$ superconformal theory
\cite{Benini:2015eyy,Benini:2016rke}
by using supersymmetry localization \cite{Benini:2015noa}.
%\cite{Benini:2015noa,Benini:2016hjo,Closset:2016arn,Honda:2015yha}.
Then there appeared agreements in various setups
involving static magnetic charged black holes \cite{Hosseini:2016tor}.
%\cite{Hosseini:2016tor,Hosseini:2016cyf,Cabo-Bizet:2017jsl,Azzurli:2017kxo,Hosseini:2017fjo,Benini:2017oxt,Halmagyi:2017hmw,Bobev:2018uxk,Hosseini:2018uzp,Crichigno:2018adf,Suh:2018tul,Hosseini:2018usu,Suh:2018szn}.

The second type of the progress has been made in the canonical AdS/CFT correspondence 
between the 4d $SU(N)$ 
$\mathcal{N}=4$ super Yang-Mills theory (SYM) and type IIB superstring theory on $AdS_5 \times S^5$,
which is also the subject of this paper.
It is known that
there are rotating electrically charged black hole solutions in $AdS_5$ \cite{Gutowski:2004ez}
%\cite{Gutowski:2004ez,Gutowski:2004yv,Chong:2005da,Chong:2005hr,Kunduri:2006ek}
which are embedded in the type IIB supergravity in $AdS_5 \times S^5$ as $1/16$-BPS solutions\footnote{
See also \cite{Colgain:2014pha} for another embedding.
} \cite{Cvetic:1999xp}. 
The black holes have three charges $(Q_1 ,Q_2 ,Q_3 )$ associated with $U(1)^3 \subset SO(6)$ 
and two angular momenta $(J_1 ,J_2 )$ associated with Cartan part of $SU(2)^2 \sim SO(4)  \subset SO(4,2)$.
They are related to the black hole mass $M$ by 
\begin{\eq}
M=g\left( |J_1 | +|J_2 | +|Q_1 | +|Q_2 | +|Q_3 | \right) ,
\end{\eq}
where $g$ is the gauge coupling.
The Bekenstein-Hawking entropy of the black hole is \cite{Kim:2006he}
\begin{\eq}
S_{\rm BH}
=\frac{\rm Area}{4G_N}
=2\pi \sqrt{Q_1 Q_2 +Q_1 Q_3 +Q_2 Q_3 -\frac{\pi}{4G_N g^3}(J_1 +J_2 )} ,
\label{eq:BH}
\end{\eq}
where the AdS/CFT dictionary between $G_N g^3$ and $N$ is 
\begin{\eq}
\frac{\pi}{2G_N g^3} =N^2 .
\label{eq:dictionary}
\end{\eq}

A long-standing question is whether 
this black hole entropy is holographically explained 
by counting $1/16$-BPS states in the $\mathcal{N}=4$ SYM on $S^1 \times S^3$.
Technically
it is much easier to analyze the superconformal index \cite{Kinney:2005ej,Romelsberger:2005eg} 
rather than the net sum of the $1/16$-BPS states:
\begin{\eq}
I_{S^1 \times S^3} 
= {\rm Tr} \Bigl[ (-1)^F e^{-\beta \{ Q,Q^\dag \}} 
p^{J_1 +\frac{r}{2}} q^{J_2 +\frac{r}{2}} v_1^{q_1} v_2^{q_2}  \Bigr] 
= {\rm Tr}_{\rm BPS} \Bigl[ (-1)^F  p^{J_1 +\frac{r}{2}} q^{J_2 +\frac{r}{2}} v_1^{q_1} v_2^{q_2}  \Bigr] ,
\end{\eq}
where $r=\frac{2}{3}(Q_1 +Q_2 +Q_3)$ and $q_{1,2}=Q_{1,2}-Q_3$
taking charges of $U(1)^3 \subset SO(6)_R$ symmetry to be $Q_{1,2,3}/2$ .
One common worry is that
the index may have huge cancellation between bosonic and fermionic states
so that it does not capture the black hole entropy \cite{Kinney:2005ej}
(see also \cite{Berkooz:2006wc} 
%\cite{Berkooz:2006wc,Janik:2007pm,Grant:2008sk,Berkooz:2008gc,Chang:2013fba} 
for other early attempts).

However,
very recent papers have updated our understanding.
First, the paper \cite{Cabo-Bizet:2018ehj} has shown that
a Legendre transformation of the black hole entropy called entropy function 
is given by a generalization of supersymmetric Casimir energy $E_{\rm Casimir}$ \cite{Assel:2014paa,Assel:2015nca} in the large-$N$ limit
which is defined as a relative factor between partition function and index\footnote{
The entropy function of the black hole was first computed in \cite{Hosseini:2017mds}.
It was also argued in \cite{Hosseini:2017mds} that
the entropy function is formally equal to the SUSY Casimir energy.
The SUSY Casimir energy of the $\mathcal{N}=4$ SYM with the fugacities of $SO(6)_R$ 
was first computed in \cite{Bobev:2015kza}.
}:
\begin{\eq}
Z_{S^1 \times S^3} =e^{-\beta E_{\rm Casimir}} I_{S^1 \times S^3} .
\label{eq:Casimir}
\end{\eq}
Second, the authors of \cite{Choi:2018hmj} have analyzed
the index of the $U(N)$ $\mathcal{N}=4$ SYM in a limit of shrinking $S^1$ at large-$N$
which we refer to as Cardy limit,
and identified a saddle point of holonomy integral which gives the black hole entropy function.
Then they have assumed the dominance of the saddle point and 
derived the asymptotic behavior of the index in the Cardy limit
which agrees with
the black hole entropy \eqref{eq:BH} via a Legendre transformation with respect to the chemical potentials.
They have also discussed a deconfinement transition in another paper \cite{Choi:2018vbz}.
Third, the authors of the paper \cite{Benini:2018ywd} have analyzed
the index for $p=q$ in the large-$N$ limit 
by using Bethe ansatz type formula of the index \cite{Benini:2018mlo}.
%\cite{Benini:2018mlo,Closset:2017bse}.
They have identified a saddle point 
which reproduces the black hole entropy function
corresponding to the equal angular momenta case: $J_1 =J_2$.
They have also assumed that the saddle point is most dominant.
It has also been stressed in \cite{Choi:2018hmj,Choi:2018vbz,Benini:2018ywd} that
the index with real fugacities have more cancellations than generic complex fugacities.

Aims of this paper are
to provide further evidence that the index gives microscopic explanation of the black hole entropy  
and make predictions for the black hole physics in more general case.
We mainly study supersymmetric index of the $SU(N)$ $\mathcal{N}=4$ SYM on $S^1 \times M_3$.
We compute an asymptotic behavior of the index 
in the limit of shrinking $S^1$ 
for arbitrary $N$
by using a refinement \cite{DiPietro:2016ond,Ardehali:2015bla}
of supersymmetric Cardy formula \cite{DiPietro:2014bca}.
Therefore our approach for the superconformal index case ($M_3 =S^3$) is basically the same as the one in \cite{Choi:2018hmj}.
The asymptotic behavior of the superconformal index at large $N$ agrees 
with the Bekenstein-Hawking entropy \eqref{eq:BH} 
via a Legendre transformation with respect to the chemical potentials.
This agreement at large-$N$ has been already found in \cite{Choi:2018hmj} recently.
We also find that the agreement formally persists for finite $N$
if we slightly modify the AdS/CFT dictionary \eqref{eq:dictionary} as
\begin{\eq}
\left. \frac{\pi}{2G_N g^3}\right|_{{\rm finite}N} =N^2 -1 =4c, 
\end{\eq}
where $c=(N^2 -1)/4$ is the central charge of the $SU(N)$ $\mathcal{N}=4$ SYM.
This implies 
an existence of non-renormalization property for the black hole entropy function in the small-$S^1$ limit at quantum level.
We also study the cases with other gauge groups and additional matters in conjugate representations,
and orbifold $\mathcal{N}=4$ SYM.
It turns out that
the entropies of all the CFT examples in this paper are given by 
\begin{\eq}
S_{\rm QFT}(Q,J) =2\pi \sqrt{Q_1 Q_2 +Q_1 Q_3 +Q_2 Q_3 -2c(J_1 +J_2 )} ,
\label{eq:S_final}
\end{\eq}
with the central charge $c$. 
This formula is our prediction for the black hole entropy with full quantum corrections. 
The results for other $M_3$ are also regarded as predictions to the gravity side\footnote{
A proposal for quantum black hole entropy for the $M_3 =S^1 \times T^2$ case 
is written in eq.~(1.82) of \cite{Hosseini:2018qsx}.
}.
It is also interesting to note that
the authors in \cite{Kim:2006he} first wrote down
the black hole formula for the dual of the $SU(N)$ $\mathcal{N}=4$ SYM as
\begin{\eq} 
S_{\rm BH}=2\pi \sqrt{Q_1 Q_2 +Q_1 Q_3 +Q_2 Q_3 -2c(J_1 +J_2 )} ,
\label{eq:BH_c}
\end{\eq}
and then substituted $c=N^2 /4$ to get the formula
\begin{\eq}
S_{\rm BH} =2\pi \sqrt{Q_1 Q_2 +Q_1 Q_3 +Q_2 Q_3 -\frac{N^2}{2} (J_1 +J_2 )} ,
\end{\eq}
in their derivation. 
Of course there is no difference in the large-$N$ limit
but our result suggests that \eqref{eq:BH_c} is more accurate for finite $N$.

Our argument for the $M_3 =S^3$ case is overlapped with the one made in \cite{Choi:2018hmj}.
While the approach is the same up to technical details and 
the final result at large-$N$ has been already obtained in \cite{Choi:2018hmj},
there are mainly three differences.
First, we mainly consider the $SU(N)$ $\mathcal{N}=4$ SYM rather than the $U(N)$ case
while the difference is irrelevant at large-$N$ and 
we also finally consider the $\mathcal{N}=4$ SYM with general gauge group as well as other theories.
Second, we analyze the index for finite $N$
but we will see that the result in \cite{Choi:2018hmj} is formally correct also for finite $N$.
Finally 
we do not only identify a saddle point giving the black hole entropy \eqref{eq:BH}
but also prove that the saddle point is most dominant.
This amounts to justify the assumption made in \cite{Choi:2018hmj} at large-$N$
and make sure that the most dominant contribution of the index gives the black hole entropy.
Some contents discussed in \cite{Choi:2018hmj} but not in this paper
are Macdonald limit \cite{Gadde:2011uv} and the case for $AdS_7$ black holes.

This paper is organized as follows.
In sec.~\ref{sec:SU}, we compute the asymptotic behavior of the SUSY index of the $SU(N)$ $\mathcal{N}=4$ SYM
in the Cardy limit $\beta\rightarrow 0$.
In sec.~\ref{sec:generalization}, 
we generalize the analysis in sec.~\ref{sec:SU} to the cases with other gauge groups and additional matters,
and the orbifold $\mathcal{N}=4$ SYM.
Sec.~\ref{sec:conclusion} is devoted to conclusion and discussions.

%%%%%%%%%%%%%%%%%%%%%%%%%%%%%%%%%%
%%%%%%%%%%%%%%%%%%%%%%%%%%%%%%%%%%
%%%%%%%%%%%%%%%%%%%%%%%%%%%%%%%%%%
\section{Asymptotic behavior of supersymmetric index in $SU(N)$ $\mathcal{N}=4$ SYM}
\label{sec:SU}
%%%%%%%%%%%%%%%%%%%%%%%%%%%%%%%%%%
%%%%%%%%%%%%%%%%%%%%%%%%%%%%%%%%%%
%%%%%%%%%%%%%%%%%%%%%%%%%%%%%%%%%%
Let us consider the $SU(N)$ $\mathcal{N}=4$ SYM
on Euclidean compact manifold of the form $S^1_\beta \times M_3$ with the radius $\beta$.
We take $M_3$ to preserve a part of supersymmetry 
and this condition constrains $S^1_\beta \times M_3$ to be complex \cite{Dumitrescu:2012ha}.
Different choices of $M_3$ count different quantum numbers
as different $M_3$'s have different isometries\footnote{
For example, an index on $T^4$ counts momenta along three ``spatial'' $S^1$'s as well as flavor charges.
}.
One of the most well-studied cases is the index on $S^1 \times S^3$ known
as superconformal index \cite{Kinney:2005ej,Romelsberger:2005eg}:
\begin{\eq}
I_{S^1 \times S^3} 
= {\rm Tr}_{\rm BPS} \Bigl[ (-1)^F  p^{J_1 +\frac{r}{2}} q^{J_2 +\frac{r}{2}} v_1^{q_1} v_2^{q_2}  \Bigr] ,
\end{\eq}
where
\begin{\eq}
p=e^{2\pi i\sigma} ,\quad
q=e^{2\pi i\tau} ,\quad
v_{1,2} =e^{2\pi i m_{1,2}} .
\label{eq:fugaticies}
\end{\eq}
We are interested in an asymptotic behavior of the partition function in the shrinking $S^1$ limit: $\beta\rightarrow 0$.
In this limit, the partition function is exactly the same as the supersymmetric index 
since we can ignore the contribution from the SUSY Casimir energy in \eqref{eq:Casimir}.
Therefore we are essentially looking at the asymptotic behavior of the index.  
There is a general formula to describe 
such asymptotic behavior for general 4d $\mathcal{N}=1$ SUSY theory with $U(1)_R$ symmetry and Lagrangian description
which is a refinement \cite{DiPietro:2016ond,Ardehali:2015bla} of SUSY Cardy formula \cite{DiPietro:2014bca}. 

For simplicity of explanations, 
we first consider the superconformal index.
We will consider more general $M_3$ later.
The superconformal index is defined through supersymmetric partition function on a space with topology of $S^1 \times S^3$.
For example, if we take $M_3$ to be the squashed sphere $S_b^3$,
$\tau$ and $\sigma$ are given by $\tau =-\beta b /2\pi i$ and $\sigma =-\beta b^{-1} /2\pi i$.
For any choices, the Cardy limit $\beta\rightarrow 0$ for the superconfomal index is equivalent to $|\tau | ,|\sigma | \rightarrow 0$.
The refined SUSY Cardy formula for the superconformal index is given by\footnote{
See \cite{Aharony:2013dha}
%\cite{Aharony:2013dha,Ardehali:2013gra,Ardehali:2013xya,DiPietro:2014bca,Golkar:2012kb,Ardehali:2014zba,Ardehali:2014esa,Ardehali:2015hya,Shaghoulian:2015kta,Shaghoulian:2015lcn,Buican:2015ina}
for earlier related works.
}
\begin{\eq}
I_{S^1 \times S^3} 
\underset{|\tau |,|\sigma | \to 0}{\simeq}
e^{-\frac{i\pi (\tau +\sigma)}{12 \tau \sigma} {\rm Tr}(R)} \int d^{{\rm rank}G} a\ 
e^{\frac{i \pi}{6\tau \sigma}V_2 (a) +\frac{i\pi  (\tau +\sigma) }{2\tau\sigma} V_1 (a)} ,
\end{\eq}
which has been derived in two ways: 
taking the limit in localization formula \cite{Ardehali:2015bla} and effective theory consideration \cite{DiPietro:2016ond}.

\begin{figure}[t]
\begin{center}
\includegraphics[clip, width=65mm]{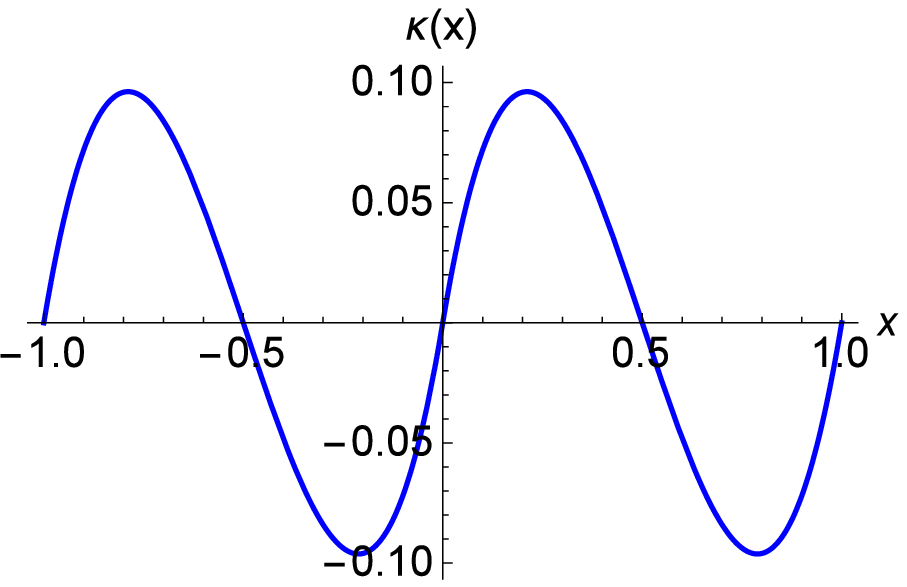}
\includegraphics[clip, width=65mm]{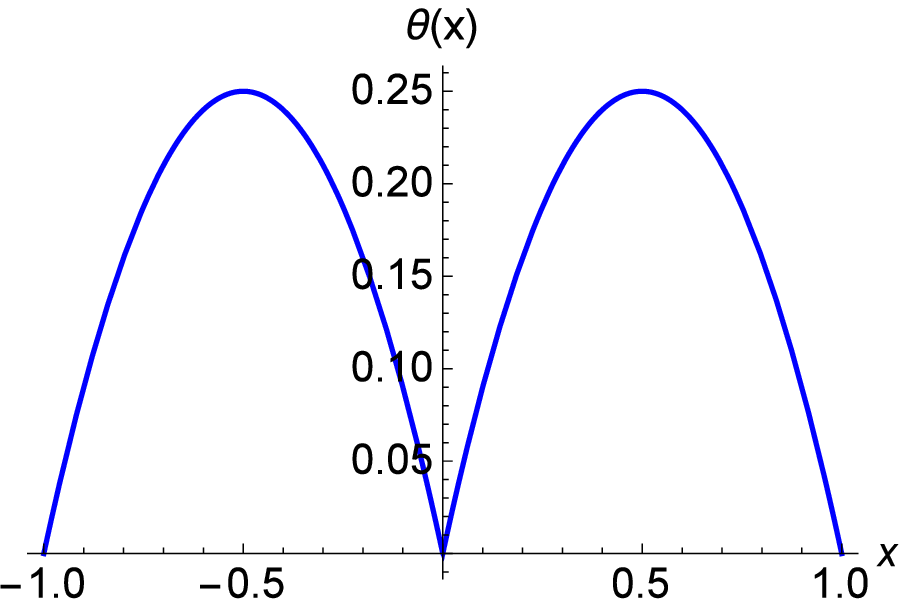}
\caption{
$\kappa (x)$ and $\theta (x)$.
}
 \label{fig:kappa}
  \end{center}
\end{figure}

Several definitions are in order.
First, $G$ is the gauge group and
$e^{2\pi i a_j}$ with $j=1,\cdots ,{\rm rank}G$ is holonomy around $S^1$ valued in the maximal torus of $G$.
Second,
${\rm Tr}(R)$ is anomaly coefficient\footnote{
This is simply the sum of $U(1)_R$ charges of fermions in theory under consideration.
} of the $U(1)_R$ symmetry
and related to conformal anomalies by ${\rm Tr}(R)=-16(c-a)$ for superconformal case.
Third,
$V_2 (a)$ and $V_1 (a)$ are piecewise polynomials of $a_j$ and flavor chemical potentials
whose forms are explicitly determined
if we specify representations, $U(1)_R$-charges and flavor charges of chiral multiplets (see app~\ref{app:general}).
Their explicit forms for the $SU(N)$ $\mathcal{N}=4$ SYM are\footnote{
For $m_1 =0=m_2$, $V_2 (a)$ and $V_1 (a)$ are zero.
The leading asymptotic behavior of the index for this case is $(N-1)\log{\beta}$ as shown in \cite{Ardehali:2015bla}.
}
\begin{\eqa}
V_2 (a) 
&=& -\sum_{1\leq i \neq j\leq N} \Bigl[ \kappa (a_{ij} +m_1 )  +\kappa (a_{ij} +m_2 ) +\kappa (a_{ij} -m_1 -m_2 ) 
   \Bigr] \NN\\
&& -(N-1) \Bigl[ \kappa (m_1 )  +\kappa (m_2 ) +\kappa (-m_1 -m_2 ) \Bigr]  ,\NN\\
V_1 (a) 
&=& \frac{1}{3} \sum_{1\leq i \neq j\leq N} \Bigl[
3\theta (a_{ij}) - \theta (a_{ij} +m_1 )  -\theta (a_{ij} +m_2 )  -\theta (a_{ij} -m_1 -m_2 )  \Bigr]  \NN\\
&& -\frac{N-1}{3} \Bigl[ \theta (m_1 )  +\theta ( m_2 )  +\theta ( -m_1 -m_2 )  \Bigr] ,
\end{\eqa}
where 
\begin{\eqa}
&& a_{ij} =a_i -a_j ,\quad  \sum_{j=1}^N a_j =0 ,\NN\\
&& \kappa (x) =\{ x\} (1 -\{ x\} )(1-2\{ x\})  ,\quad 
\theta (x) =\{x \} (1-\{ x\}) ,
\end{\eqa}
with fractional part $\{ x\} \equiv x -[x]$
(see fig.~\ref{fig:kappa} for shapes of $\kappa (x)$ and $\theta (x)$). 
$V_2 (a)$ ($V_1 (a)$) is apparently a piecewise cubic (quadratic) polynomial
but this is actually quadratic (linear) 
because there is a cancellation of the highest order terms 
physically coming from cancellation of anomalies involving the gauge symmetry.

%%%%%%%%%%%%%%%%%%%%%%
%%%%%%%%%%%%%%%%%%%%

\begin{figure}[t]
\begin{center}
\includegraphics[clip, width=65mm]{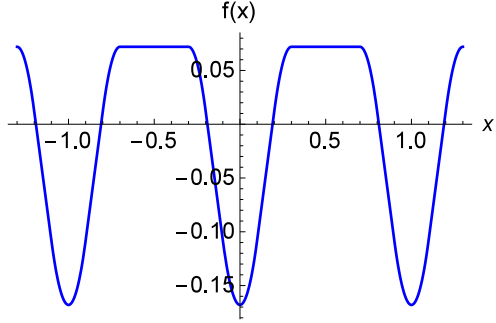}
\caption{
$f(x)$ for $(m_1 ,m_2 )=(0.2, 0.1)$.
}
 \label{fig:fx}
  \end{center}
\end{figure}

Here we restrict ourselves to
\begin{\eq}
{\rm Re}\left( \frac{i}{\tau\sigma} \right) < 0 , 
\label{eq:tau_sigma}
\end{\eq}
and mention other regime later.
In this regime,
the integral in the limit is dominated by saddle point configuration(s) 
to minimize the function $V_2 (a)$.
We can easily find a dominant saddle point as follows.
Noting $\kappa (-x)=-\kappa (x)$ and\footnote{
Physically this periodicity reflects invariance under large gauge transformation.
} $\kappa (x+1)=\kappa (x)$, we rewrite $V_2 (a)$ as
\begin{\eq}
V_2 (a)= \sum_{i<j} f(a_{ij }) +\frac{N-1}{2}f(0) ,
\end{\eq}
where
\begin{\eqa}
f(a_{ij})
&=&   \kappa (a_{ij} -\{ m_1 \}  ) -\kappa (a_{ij} +\{ m_1 \}  )
   +\kappa (a_{ij} -\{ m_2 \} ) -\kappa (a_{ij} +\{ m_2 \}  )  \NN\\
&&   +\kappa (a_{ij} +\{m_1 \} +\{ m_2 \}  ) -\kappa (a_{ij} - \{ m_1 \} -\{ m_2 \}  )  .
\end{\eqa}
It is sufficient to minimize each $f(a_{ij})$ 
and show that we can realize a simultaneously minimizing configuration.
As a result, the minimizing configuration is simply $a_j =0$ for any $j$
as illustrated in fig.~\ref{fig:fx} for specific values of $(m_1 ,m_2 )$.
To see this generally,
it is convenient to first analyze the regime
\begin{\eqa}
 0\leq  \{ m_2 \} \leq \{ m_1 \}  ,\quad  \{ m_1 \} +\{ m_2 \} \leq \frac{1}{2}  ,
\end{\eqa}
and extend it to other regime by using the periodicity $m_{1,2} \sim m_{1,2} +1$.
In this regime,
noting $\kappa (x)=2x^3 -3x|x| +x$ for $|x|\leq 1$,
the function $f(x)$ in ``the fundamental region'' $|x|< 1- \{ m_1 \} +\{m_2 \}$ is given by
\begin{\eq}
f(x) 
=\begin{cases}
6 x^2 +12 \{ m_1 \} \{ m_2 \}  ( \{ m_1 \} +\{ m_2 \} -1 ) 
& {\rm for}\ |x|\leq \{ m_2 \} \cr
12m_2 |x| +6m_2  (2 m_1^2  +2m_1 m_2  -2m_1  -m_2 )
 & {\rm for}\ \{ m_2 \} \leq  |x| \leq \{ m_1 \} \cr
-6 (|x| -\{ m_1 \} -\{ m_2 \} )^2  +12\{ m_1 \} \{ m_2 \}  (\{ m_1 \} +\{ m_2 \} )
& {\rm for}\ \{ m_1 \} \leq  |x| \leq \{ m_1 \} +\{ m_2 \} \cr
12 \{ m_1 \} \{ m_2 \}  (\{ m_1 \} +\{ m_2 \} )  
& {\rm for}\ \{ m_1 \} +\{ m_2 \} \leq |x|  
\end{cases} ,
\label{eq:fx}
\end{\eq}
which has the minimum at the origin:
\begin{\eq}
\left. f(x) \right|_{\rm min} = f(0) = 12 \{ m_1 \} \{ m_2 \} ( \{ m_1 \} + \{ m_2 \} -1) .
\end{\eq}
Therefore the minimum of $V_2 (a)$ is realized 
by $a_{ij}=0$ for all $i,j$ with the traceless condition $\sum_{j=1}^N a_j =0$,
which is nothing but $a_j =0$.
Thus we find the minimum of $V_2 (a)$ as 
\begin{\eq}
\left. V_2 (a) \right|_{\rm min} =V_2 (0) =6 (N^2 -1) \{ m_1 \} \{ m_2 \} (\{ m_1 \} +\{ m_2 \} -1) .
\end{\eq}

%%%%%%%%%%%%%%%%%%%%%%
%%%%%%%%%%%%%%%%%%%%

The next order $\mathcal{O}(\beta^{-1})$ is simply obtained by substituting\footnote{
The saddle point of $V_2 (a)$ also realizes the minimum of $V_1 (a)$ as a result 
though this property is not necessary for our analysis.
Beyond this order, we need to take into account fluctuations around the saddle point.
} 
the saddle point into $V_1 (a)$:
\begin{\eq}
\left. V_1 (a)  \right|_{a_j =0} 
=\frac{2 (N^2 -1 )}{3} \Bigl[ 
 \{ m_1 \}^2 +\{ m_2 \}^2 +\{m_1 \} \{ m_2 \}  -\{ m_1 \}  - \{ m_2 \}
 \Bigr] .
\end{\eq}
Then, noting $c-a =0$ in the $\mathcal{N}=4$ SYM,
we find the Cardy limit of the superconformal index to be
\begin{\eqa}
\log{I_{S^1 \times S^3}} 
&\underset{|\tau |,|\sigma | \to 0}{\simeq} &
\frac{i\pi (N^2 -1)}{\tau \sigma} \Biggl[ 
 \{ m_1 \} \{ m_2 \} (\{ m_1 \} +\{ m_2 \} -1) \NN\\
&& 
+\frac{\tau +\sigma }{3} \left(  \{ m_1 \}^2 +\{ m_2 \}^2 +\{m_1 \} \{ m_2 \}  -\{ m_1 \}  - \{ m_2 \}  \right)
 \Bigr] .
\end{\eqa}
In order to directly compare this with the Bekenstein-Hawking entropy,
it is convenient to rewrite the result in the following two steps.
First we redefine the chemical potentials $m_{1,2}$ as
\begin{\eq}
m_{1,2} = \Delta_{1,2} -\frac{\tau +\sigma}{3} ,
\end{\eq}
so that our index becomes
\begin{\eq}
{\rm Tr}_{\rm BPS} \Bigl[ (-1)^F  p^{J_1 +Q_3 } q^{J_2 +Q_3 } 
e^{2\pi i \Delta_1 (Q_1 -Q_3 ) }  e^{2\pi i \Delta_2 (Q_2 -Q_3 ) } \Bigr] .
\end{\eq}
This object is the same as the grand canonical partition function
\begin{\eq}
{\rm Tr}_{\rm BPS} 
\Bigl[ p^{J_1 } q^{J_2 } \prod_{a=1}^3 e^{2\pi i \Delta_a Q_a } \Bigr] ,
\end{\eq}
with the constraint\footnote{
We have used $(-1)^F =e^{2\pi i Q_3}$.
} $\Delta_1 +\Delta_2  +\Delta_3 -\tau -\sigma -1 \in 2\mathbb{Z}$.
In this parametrization, the asymptotic behavior of the index becomes
\begin{\eq}
\log{I_{S^1 \times S^3}} 
\underset{|\tau |,|\sigma | \to 0}{\simeq}
\frac{i\pi (N^2 -1) \{ \Delta_1 \} \{ \Delta_2 \} (\{ \Delta_1 \} +\{ \Delta_2 \}-1 -\sigma -\tau  ) }{\tau\sigma} .
\end{\eq}
Second, we perform a Legendre transformation \cite{Hosseini:2017mds}
with respect to $(\sigma ,\tau , \Delta_1 ,\Delta_2 )$
to directly obtain entropy 
or equivalently degeneracy of states with fixed charges and angular momenta. 
We will perform this analysis in next subsection.

%%%%%%%%%%%%%%%%%%%%%%%%%%%%%%%%%%
\subsubsection*{Comments on other regime of $(\tau ,\sigma )$}
%%%%%%%%%%%%%%%%%%%%%%%%%%%%%%%%%%
So far we have taken ${\rm Re}\left( \frac{i}{\tau\sigma} \right) < 0$.
If we take it oppositely i.e.~${\rm Re}\left( \frac{i}{\tau\sigma} \right) > 0$,
then we need to minimize $-V_2 (a)$ or equivalently maximize $V_2 (a)$.
Then the dominant saddle points are given by the points maximizing $f(x)$.
According to \eqref{eq:fx},
the saddle points are any configurations giving the plateau regime of $f(x)$,
namely the ones satisfying $\{ m_1 \} +\{ m_2 \} \leq  | \{ a_{ij} \}| < 1- \{ m_1 \} +\{m_2 \}$.
We immediately see that the saddle points are no longer isolated and 
therefore it remains integration over the saddle points
which seems complicated since $V_1 (a)$ is not constant in this regime. 
As a result, the asymptotic behavior of the index is
\begin{\eqa}
\log{I_{S^1 \times S^3}} 
&\underset{|\tau |,|\sigma | \to 0}{\simeq} & 
\frac{i\pi \{ m_1 \} \{ m_2 \}}{\tau \sigma} \Biggl[ 
 (N^2 -1) (\{ m_1 \} +\{ m_2 \} ) -(N-1) \Bigr] \NN\\
&& +\log \int_{\rm saddles} d^N a\ \delta \Bigl(\sum_{j=1}^N a_j \Bigr)  
e^{\frac{i\pi (\tau +\sigma)}{2\tau \sigma}V_1 (a)} .
\end{\eqa}
This implies that
we have anti-Stokes line at ${\rm Re}\left( \frac{i}{\tau\sigma} \right) = 0$
since the dominant saddle point changes there.
The above saddle points are unstable
in the regime ${\rm Re}\left( \frac{i}{\tau\sigma} \right) < 0$ which we have mainly considered in this paper.
Relatedly
Stokes phenomena have been observed in the large-$N$ analysis of the Bethe ansatz type formula \cite{Benini:2018ywd}.
It is interesting to understand the above phenomena in more detail
and find their physical interpretations especially from the gravity side.
This might be related to hairy black holes discussed in \cite{Bhattacharyya:2010yg}.
%\cite{Bhattacharyya:2010yg,Dias:2011tj,Markeviciute:2016ivy,Markeviciute:2018yal,Markeviciute:2018cqs}.

%%%%%%%%%%%%%%%%%%%%%%%%%%%%%%%%%%
%%%%%%%%%%%%%%%%%%%%%%%%%%%%%%%%%%
\subsection{Comparison with Bekenstein-Hawking entropy}
%%%%%%%%%%%%%%%%%%%%%%%%%%%%%%%%%%
%%%%%%%%%%%%%%%%%%%%%%%%%%%%%%%%%%
This subsection is essentially a review of 
various papers \cite{Hosseini:2017mds,Cabo-Bizet:2018ehj,Choi:2018hmj,Benini:2018ywd}
up to identifications of parameters and final results\footnote{
The original argument was in sec.~3 of \cite{Hosseini:2017mds}.
This subsection is also a review of appendix B of \cite{Cabo-Bizet:2018ehj},
sec.~2.3 of \cite{Choi:2018hmj} and sec.~6 of \cite{Benini:2018ywd}.
See also \cite{Hosseini:2018dob} for a similar argument for $AdS_7$ black holes.
}.
The Legendre transformation of the black hole entropy is referred to as entropy function \cite{Sen:2005wa}.
Suppose that we have the entropy function $\mathcal{S}$:
\begin{\eq}
\mathcal{S}
= 2\pi i \nu \frac{X_1 X_2 X_3 }{\omega_1 \omega_2}  ,
\end{\eq}
with the constraint
\begin{\eq}
X_1 +X_2 +X_3 -\omega_1 -\omega_2 = n.
\label{eq_constraint}
\end{\eq}
These quantities in our case are
\begin{\eq}
\mathcal{S} =-\log{I_{S^1 \times S^3_b}}  ,\quad  \nu =\frac{N^2 -1}{2},\quad 
\omega_1 =\sigma, \quad \omega_2 =\tau ,\quad
X_a =\{ \Delta_a \} ,\quad n=1 .
\label{eq:identify}
\end{\eq}
The entropy $S(Q,J)$ is obtained by the Legendre transformation
\begin{\eq}
S(Q,J)
=\left. \mathcal{S}(X_a , \omega_i )
 +2\pi i \left( \sum_{a=1}^3 X_a  Q_a +\sum_{I=1}^2 \omega_I J_I \right) 
 +2\pi i \Lambda  \left( \sum_{a=1}^3 X_a -\sum_{I=1}^2 \omega_I -n \right)
\right|_{X_a ,\omega_i } ,
\end{\eq}
where $\Lambda$ is Lagrange multiplier.
The extremization conditions are
\begin{\eq}
\frac{\del \mathcal{S}}{\del X_a} =-2\pi i (Q_a +\Lambda ), \quad
\frac{\del \mathcal{S}}{\del \omega_I } =-2\pi i (J_I  -\Lambda ) ,
\end{\eq}
with the constraint \eqref{eq_constraint}.
Note that we do not need explicit solutions for $(X_a ,\omega_I )$ to compute $S$
if we use the relation 
\begin{\eq}
\mathcal{S} = \sum_{a=1}^3 X_a \frac{\del \mathcal{S}}{\del X_a } +\sum_{I=1}^2 \omega_I \frac{\del \mathcal{S}}{\del \omega_I} .
\end{\eq}
Then the entropy is simply given by
\begin{\eq}
S= 2\pi in \Lambda ,
\end{\eq}
where $\Lambda$ satisfies
\begin{\eq}
 0=(Q_1 +\Lambda )(Q_2 +\Lambda )(Q_3 +\Lambda ) +\nu (J_1 -\Lambda )(J_2 -\Lambda ) 
=\Lambda^3 +p_2 \Lambda^2 +p_1 \Lambda +p_0 , 
\end{\eq}
with
\begin{\eqa}
&& p_0 
= Q_1 Q_2 Q_3 +\mu J_1 J_2 ,\NN\\
&& p_1 
= Q_1 Q_2 +Q_2 Q_3 +Q_3 Q_1 -\nu  (J_1 +J_2 ) ,\NN\\
&& p_2 
= Q_1 +Q_2 +Q_3 +\mu J_1 J_2 .
\end{\eqa}
The equation for $\Lambda$ has the three solutions $\Lambda = \{ -p_2 ,\pm i \sqrt{p_1} \}$ with $p_1 ,p_2 \in\mathbb{R}_{\geq 0}$.
Imposing the entropy to be real positive,
the physical solution among the three is $\Lambda = -i {\rm sign}(n) \sqrt{p_1}$.
which leads us to the entropy
\begin{\eq}
S=2\pi |n| \sqrt{p_1} .
\end{\eq}
Under the identifications \eqref{eq:identify},
the entropy computed by the superconformal index of the $SU(N)$ $\mathcal{N}=4$ SYM is
\begin{\eq}
S_{\rm QFT}(Q,J) =2\pi \sqrt{Q_1 Q_2 +Q_1 Q_3 +Q_2 Q_3 -\frac{N^2 -1}{2}(J_1 +J_2 )} ,
\end{\eq}
which agrees with the Bekenstein-Hawking entropy \eqref{eq:BH}
via the AdS/CFT dictionary \eqref{eq:dictionary} in the large-$N$ limit.
Interestingly,
the agreement persists for finite $N$
if we slightly modify 
the AdS/CFT dictionary for finite $N$ as
\begin{\eq}
\left. \frac{\pi}{2G_N g^3}\right|_{{\rm finite}N} =N^2 -1 =4c,
\label{eq:Newton}
\end{\eq}
where $c=\frac{N^2 -1}{4}$ is the central charge.
This may suggest that
the black hole entropy with full quantum corrections
is captured by the Bekenstein-Hawking entropy with the renormalized Newton constant \eqref{eq:Newton} in the Cardy limit.

%%%%%%%%%%%%%%%%%%%%%%%%%%%%%%%%%%
%%%%%%%%%%%%%%%%%%%%%%%%%%%%%%%%%%
\subsection{General $M_3$}
%%%%%%%%%%%%%%%%%%%%%%%%%%%%%%%%%%
%%%%%%%%%%%%%%%%%%%%%%%%%%%%%%%%%%
The refined SUSY Cardy formula for the $SU(N)$ $\mathcal{N}=4$ SYM on $S^1 \times M_3$ is 
\begin{\eq}
I_{S^1 \times M_3}
\underset{\beta \to 0}{\simeq} 
\int d^N a\ \delta \left( \sum_{j=1}^N a_j \right) 
e^{-\frac{\pi^3 i A_{M_3}}{6\beta^2}V_2 (a) +\frac{\pi^2 L_{M_3} }{2\beta} V_1 (a)  -\frac{1}{2\beta} \tilde{V}_1 (a) }
\end{\eq}
where $\tilde{V}_1 (a)$ is the contribution absent in the superconformal index:
\begin{\eq}
\tilde{V}_1 (a)
= \sum_{i\neq j} ( \ell_{M_3}^i -\ell_{M_3}^j ) \Bigl[
\theta (a_{ij}+m_1 ) +\theta (a_{ij}+m_2 ) +\theta (a_{ij} -m_1 -m_2 ) +\theta (a_{ij} )
\Bigr] .
\end{\eq}
The quantities $A_{M_3}$, $L_{M_3}$ and $\ell^{i}_{M_3}$ are local functionals on $M_3$
given by bosonic fields in the 3d new minimal supergravity multiplet $(h_{\mu\nu} ,A_\mu^{(R)},H,c_\mu )$
and 3d $\mathcal{N}=2$ vector multiplet\footnote{
This is both for gauge and global symmetries.
} $(A_\mu ,\sigma ,D )$:
\begin{\eqa}
A_{M_3} 
&=& \frac{i}{\pi^2}\int_{M_3}d^3 x \sqrt{h}  \Bigl[-c^\mu v_\mu+ 2 H \Bigr] , \NN\\
L_{M_3} 
&=& \frac{1}{ \pi^2}\int_{M_3}d^3 x \sqrt{h}  \Bigl[ -A_\mu^{(R)\mu} v^\mu  +  v_\mu v^\mu - \frac{1}{2} H^2  +  \frac{1}{4} R \Bigr] ,\NN\\ 
\ell^{i}_{M_3} 
& =& \frac{1}{\pi^2}\int_{M_3}d^3 x \sqrt{h} \Bigl[-A_\mu^i v^\mu +  D^i \Bigr] ,
\end{\eqa}
which come from induced Chern-Simons terms of $U(1)_{\rm KK}$-$U(1)_{\rm KK}$,
$U(1)_{\rm KK}$-$U(1)_R$ and $U(1)_{\rm KK}$-Gauge/Flavor respectively,
from the viewpoint of 3d effective theory\footnote{
See \cite{DiPietro:2014bca,DiPietro:2016ond} for details.
} on $M_3$.
Technically $A_{M_3}$ and $L_{M_3}$ are just constants for fixed $M_3$
while $l^i_{M_3}$ generally depends on (supersymmetric configurations of) the dynamical vector multiplets
though it has typically a simple form because of SUSY\footnote{
For example, $\ell^i_{S_3 /\mathbb{Z}_n} =0$ 
and $\ell_{S^1 \times \Sigma_g}^i \propto ({\rm magnetic\ charge})$ 
with Riemann surface $\Sigma_g$.
}.

Here we restrict ourselves to 
\begin{\eq}
{\rm Re}\left( \frac{i A_{M_3}}{\beta^2} \right) >0 ,
\end{\eq}
which generalizes the condition \eqref{eq:tau_sigma}.
Then the integral in the $\beta\rightarrow 0$ limit is dominated 
by the saddle point of $V_2 (a)$ which is already found as $a_j =0$.
Thus,
noting $\left. \tilde{V}_1 (a) \right|_{a_j =0} =0$,
the asymptotic behavior of the index for general $M_3$ is 
\begin{\eqa}
\log{I_{S^1 \times M_3}} 
&\underset{\beta \to 0}{\simeq}&
 -\frac{2\pi^3 i A_{M_3} (N^2 -1)}{\beta^2}   \{ m_1 \} \{ m_2 \} ( \{ m_1 \} + \{ m_2 \} -1) \NN\\
&& +\frac{\pi^2 L_{M_3} (N^2 -1 )}{3\beta} \Bigl[ 
 \{ m_1 \}^2 +\{ m_2 \}^2 +\{m_1 \} \{ m_2 \}  -\{ m_1 \}  - \{ m_2 \}
 \Bigr]  .\NN\\
\end{\eqa}
This makes predictions to the gravity side for more general $M_3$.
For example,
the case for Lens space index is
\begin{\eqa}
\log{I_{S^1 \times S^3/\mathbb{Z}_n}} 
&\underset{|\tau |,|\sigma | \to 0}{\simeq} & 
\frac{i\pi (N^2 -1)}{n\tau \sigma} \Biggl[ 
 \{ m_1 \} \{ m_2 \} (\{ m_1 \} +\{ m_2 \} -1) \NN\\
&& 
+\frac{\tau +\sigma }{3} \left(  \{ m_1 \}^2 +\{ m_2 \}^2 +\{m_1 \} \{ m_2 \}  -\{ m_1 \}  - \{ m_2 \}  \right)
 \Bigr]  \NN\\
&=& \frac{\log{I_{S^1 \times S^3}}}{n} ,
 \end{\eqa}
which implies that the dual black hole entropy is $1/n$ of the one for the superconformal index.

%%%%%%%%%%%%%%%%%%%%%%%%%%%%%%%%%%
%%%%%%%%%%%%%%%%%%%%%%%%%%%%%%%%%%
%%%%%%%%%%%%%%%%%%%%%%%%%%%%%%%%%%
\section{Generalizations}
\label{sec:generalization}
%%%%%%%%%%%%%%%%%%%%%%%%%%%%%%%%%%
%%%%%%%%%%%%%%%%%%%%%%%%%%%%%%%%%%
%%%%%%%%%%%%%%%%%%%%%%%%%%%%%%%%%%
%%%%%%%%%%%%%%%%%%%%%%%%%%%%%%%%%%
%%%%%%%%%%%%%%%%%%%%%%%%%%%%%%%%%%
\subsection{Other gauge groups}
%%%%%%%%%%%%%%%%%%%%%%%%%%%%%%%%%%
%%%%%%%%%%%%%%%%%%%%%%%%%%%%%%%%%%
Generalization to other gauge groups is straightforward
because we can still apply the technique in the $SU(N)$ case.
For the $\mathcal{N}=4$ SYM with gauge group $G$,
the functions appearing in the SUSY Cardy formula are
\begin{\eqa}
V_2 (a) 
&=& -\sum_{\alpha\in {\rm root}} \Bigl[ \kappa ( \alpha\cdot a +m_1 )  +\kappa (\alpha\cdot a +m_2 ) +\kappa ( \alpha\cdot a -m_1 -m_2 ) 
   \Bigr] \NN\\
&& -{\rm rank}(G)  \Bigl[ \kappa (m_1 )  +\kappa (m_2 ) +\kappa (-m_1 -m_2 ) \Bigr]  ,\NN\\
V_1 (a) 
&=& \frac{1}{3} \sum_{\alpha\in {\rm root}} \Bigl[ 
3\theta (\alpha\cdot a) -\theta (\alpha\cdot a +m_1 )  -\theta (\alpha\cdot a +m_2 )  -\theta (\alpha\cdot a -m_1 -m_2 )  \Bigr]  \NN\\
&& -\frac{N-1}{3} \Bigl[ \theta (m_1 )  +\theta ( m_2 )  +\theta ( -m_1 -m_2 )  \Bigr] ,\NN\\
\tilde{V}_1 (a)
&=& \sum_{\alpha\in {\rm root}} \alpha\cdot \ell_{M_3} \Bigl[
\theta (\alpha\cdot a +m_1 ) +\theta (\alpha\cdot a  +m_2 ) +\theta (\alpha\cdot a  -m_1 -m_2 ) +\theta (\alpha\cdot a  )
\Bigr] . \NN\\
\end{\eqa}
In terms of $f(x)$, we rewrite $V_2 (a)$ as
\begin{\eq}
V_2 (a)= \sum_{\alpha \in {\rm root}_+} f(\alpha\cdot a ) +\frac{{\rm rank}(G)}{2}f(0) ,
\end{\eq}
which has the global minimum at $a_j =0$ by the same logic\footnote{
For $G=U(N)$, this is sufficient but not necessary due to decoupling the diagonal $U(1)$.
The same minimum is realized by any configuration satisfying $a_1 = \cdots =a_N$
which is the same as the one obtained in \cite{Choi:2018hmj}.
This flat direction affects $\mathcal{O}(\log{\beta} )$.
} as in sec.~\ref{sec:SU}.
Thus the index asymptotically behaves as
\begin{\eqa}
\log{I_{S^1 \times M_3}} 
&\underset{\beta \to 0}{\simeq}&
 -\frac{2\pi^3 i A_{M_3}{\rm dim}(G)}{\beta^2}   \{ m_1 \} \{ m_2 \} ( \{ m_1 \} + \{ m_2 \} -1) \NN\\
&& +\frac{\pi^2 L_{M_3} {\rm dim}(G)}{3\beta} \Bigl[ 
 \{ m_1 \}^2 +\{ m_2 \}^2 +\{m_1 \} \{ m_2 \}  -\{ m_1 \}  - \{ m_2 \}
 \Bigr]  .
\end{\eqa}
Especially, the superconformal index is\footnote{
For $G=U(N)$, the result is the same as the one obtained in \cite{Choi:2018hmj}
which takes the large-$N$ limit.
However, our result shows that the result of \cite{Choi:2018hmj} is formally correct also for finite $N$.
This implies that contributions which are ignored in \cite{Choi:2018hmj} vanish in the Cardy limit.
}
\begin{\eq}
\log{I_{S^1 \times S^3}} 
\underset{|\tau |,|\sigma | \to 0}{\simeq}
\frac{i\pi {\rm dim}G \{ \Delta_1 \} \{ \Delta_2 \} (\{ \Delta_1 \} +\{ \Delta_2 \}-1 -\sigma -\tau  ) }{\tau\sigma} .
\end{\eq}
The Legendre transformation leads us to the entropy
\begin{\eqa}
S_{\rm QFT}(Q,J) 
&=& 2\pi \sqrt{Q_1 Q_2 +Q_1 Q_3 +Q_2 Q_3 -\frac{{\rm dim}G }{2}(J_1 +J_2 )}  \NN\\
&=& 2\pi \sqrt{Q_1 Q_2 +Q_1 Q_3 +Q_2 Q_3 -2c(J_1 +J_2 )} ,
\end{\eqa}
where we have used $c={\rm dim}G /4$.
This implies that
the dual black hole entropy for gauge group $G$ 
is captured by \eqref{eq:BH}
under the identification
\begin{\eq}
\left. \frac{\pi}{2G_N g^3}\right|_{{\rm finite}N} =4c ,
\end{\eq}
even if $G$ is not necessarily $SU(N)$ or $U(N)$.

%%%%%%%%%%%%%%%%%%%%%%%%%%%%%%%%%%
%%%%%%%%%%%%%%%%%%%%%%%%%%%%%%%%%%
\subsection{Adding matters in conjugate representations}
%%%%%%%%%%%%%%%%%%%%%%%%%%%%%%%%%%
%%%%%%%%%%%%%%%%%%%%%%%%%%%%%%%%%%
Let us add pairs of chiral multiplets in conjugate representations 
to the $\mathcal{N}=4$ SYM with the gauge group $G$.
In general this theory may have new flavor symmetries
but let us keep to turn off fugacities of the new symmetries for simplicity.
For this case,
the function $V_2 (a)$ does not receive contributions from the additional matters
essentially because of $\kappa (-x) =-\kappa (x)$.
Therefore the holonomy integral of the SUSY Cardy formula is still dominated by $a_j =0$.
Furthermore,
contributions from the additional matters to the $V_1 (a)$ and $\tilde{V}_1 (a)$
are zero at $a_j =0$.
Thus,
the asymptotic behavior of the index is
\begin{\eqa}
\log{I_{S^1 \times M_3}} 
&\underset{\beta \to 0}{\simeq}&
-\frac{\pi^2 L_{M_3}}{12\beta} {\rm Tr}(R)
-\frac{2\pi^3 i A_{M_3}{\rm dim}(G)}{\beta^2}   \{ m_1 \} \{ m_2 \} ( \{ m_1 \} + \{ m_2 \} -1) \NN\\
&& +\frac{\pi^2 L_{M_3} {\rm dim}(G)}{3\beta} \Bigl[ 
 \{ m_1 \}^2 +\{ m_2 \}^2 +\{m_1 \} \{ m_2 \}  -\{ m_1 \}  - \{ m_2 \}
 \Bigr] . 
\end{\eqa}
Note that the difference from the $\mathcal{N}=4$ SYM is only the first term,
which is simply captured by the unrefined SUSY Cardy formula \cite{DiPietro:2014bca}.
Specifying to the superconformal index case, we find
\begin{\eq}
\log{I_{S^1 \times S^3}} 
\underset{|\tau |,|\sigma | \to 0}{\simeq}
 \frac{i\pi {\rm dim}G \{ \Delta_1 \} \{ \Delta_2 \} (\{ \Delta_1 \} +\{ \Delta_2 \}-1 -\sigma -\tau  ) }{\tau\sigma}
 -\frac{i\pi (\tau +\sigma )}{12\tau\sigma}{\rm Tr}(R) .
\end{\eq}
This indicates that
the entropies in theories with $|{\rm Tr}(R)|/N^2 \ll 1$ in the large-$N$ limit are universally captured
by the one of the $\mathcal{N}=4$ SYM.
An interesting example of such theories is
the $SU(N)$ $\mathcal{N}=4$ SYM plus $N_f$ fundamental hypermultiples known as D3-D7 system.

%%%%%%%%%%%%%%%%%%%%%%%%%%%%%%%%%%
%%%%%%%%%%%%%%%%%%%%%%%%%%%%%%%%%%
\subsection{Orbifold $\mathcal{N}=4$ SYM}
%%%%%%%%%%%%%%%%%%%%%%%%%%%%%%%%%%
%%%%%%%%%%%%%%%%%%%%%%%%%%%%%%%%%%
Let us consider so-called orbifold $\mathcal{N}=4$ SYM
which is the circular quiver $\mathcal{N}=2$ gauge theory 
with $U(N)_1 \times \cdots U(N)_K$ gauge group
and one bi-fundamental hypermultiplet of neighboring gauge group\footnote{
in the notation $U(N)_{K+1}= U(N)_{1}$.
} $U(N)_I \times U(N)_{I+1}$.
We turn on chemical potentials $m_1 ,m_2$ of flavor symmetry $U(1)_1 \times U(1)_2$
in which
the $U(1)_1$  ($U(1)_2$) symmetry assigns
charge 1 to each $\mathcal{N}=1$ (anti-)bi-fundamental chiral multiplet
and charge -1 to each $\mathcal{N}=1$ adjoint chiral multiplet in the $\mathcal{N}=2$ vector multiplet.
The function $V_2 (a)$ for this theory is
\begin{\eq}
V_2 (a)
= -\sum_{I=1}^K \sum_{1\leq i,j \leq N} \Biggl[ 
    \kappa \left( a_i^{(I)} -a_j^{(I+1)} +m_1 \right) 
+\kappa \left( -a_i^{(I)} +a_j^{(I+1)} +m_2 \right)   
 + \kappa \left( a_{ij}^{(I)}  -m_1-m_2 \right) 
\Biggr] .
\end{\eq}
It is not easy to find global minimum of this function in contrast to the $\mathcal{N}=4$ SYM.
Instead of solving this problem completely,
we proceed by taking the physically motivated ansatz:
\begin{\eq}
a_j^{(I)} =a_j^{(J)} = a_j ,
\end{\eq}
which reflects $\mathbb{Z}_k$ rotation symmetry of the quiver diagram 
or equivalently all the gauge groups are ``democratic"\footnote{
This type of ansatz was taken also in large-$N$ analysis of $S^4$ partition function in the orbifold $\mathcal{N}=4$ SYM \cite{Azeyanagi:2013fla}.
}.
Under this ansatz, $V_2 (a)$ becomes 
\begin{\eq}
\left. V_2 (a) \right|_{a_j^{(I)} =a_j^{(J)} = a_j}
= - K \sum_{1\leq i,j \leq N} \Bigl[ 
    \kappa \left( a_{ij} +m_1 \right) +\kappa \left( -a_{ij} +m_2 \right)    + \kappa \left( a_{ij}  -m_1-m_2 \right) 
\Bigr] ,
\end{\eq}
which is proportional to $V_2 (a)$ of the $U(N)$ $\mathcal{N}=4$ SYM.
Thus, the asymptotic behavior of the index is
\begin{\eqa}
\log{I_{S^1 \times M_3}} 
&\underset{\beta \to 0}{\simeq}&
 -\frac{2\pi^3 i A_{M_3} KN^2}{\beta^2}   \{ m_1 \} \{ m_2 \} ( \{ m_1 \} + \{ m_2 \} -1) \NN\\
&& +\frac{\pi^2 L_{M_3} KN^2 }{3\beta} \Bigl[ 
 \{ m_1 \}^2 +\{ m_2 \}^2 +\{m_1 \} \{ m_2 \}  -\{ m_1 \}  - \{ m_2 \}
 \Bigr]  .
\end{\eqa}
This result has a nice interpretation from the viewpoint of so-called large-$N$ orbifold equivalence \cite{Kachru:1998ys}
%\cite{Kachru:1998ys,Bershadsky:1998cb,Kovtun:2004bz}
which states that 
a free energy of a ``daughter" theory obtained by a projection of a ``parent" theory by a group $\Gamma$ obeys
\begin{\eq}
\lim_{N\rightarrow\infty } \frac{F_{\rm daughter}}{N^2} 
=\frac{1}{|\Gamma |} \lim_{N\rightarrow\infty }  \frac{F_{\rm parent}}{ N^2} ,
\end{\eq}
where $|\Gamma |$ is the order of $\Gamma$.
Since the orbifold $\mathcal{N}=4$ SYM is obtained by a $\mathbb{Z}_K$ projection of the $U(KN)$ $\mathcal{N}=4$ SYM,
the above result is expected from the orbifold equivalence.
The result for the superconformal index is
\begin{\eq}
\log{I_{S^1 \times S^3}} 
\underset{|\tau |,|\sigma | \to 0}{\simeq}
\frac{i\pi KN^2 \{ \Delta_1 \} \{ \Delta_2 \} (\{ \Delta_1 \} +\{ \Delta_2 \}-1 -\sigma -\tau  ) }{\tau\sigma} ,
\end{\eq}
which gives the entropy
\begin{\eq}
S_{\rm QFT}(Q,J) =2\pi \sqrt{Q_1 Q_2 +Q_1 Q_3 +Q_2 Q_3 -\frac{KN^2}{2}(J_1 +J_2 )} .
\end{\eq}
Noting $c=KN^2 /2$, we can also express this as
\begin{\eq}
S_{\rm QFT}(Q,J) =2\pi \sqrt{Q_1 Q_2 +Q_1 Q_3 +Q_2 Q_3 -2c(J_1 +J_2 )} .
\end{\eq}

%%%%%%%%%%%%%%%%%%%%%%%%%%%%%%%%%%
%%%%%%%%%%%%%%%%%%%%%%%%%%%%%%%%%%
%%%%%%%%%%%%%%%%%%%%%%%%%%%%%%%%%%
\section{Conclusion and Discussions}
\label{sec:conclusion}
%%%%%%%%%%%%%%%%%%%%%%%%%%%%%%%%%%
%%%%%%%%%%%%%%%%%%%%%%%%%%%%%%%%%%
%%%%%%%%%%%%%%%%%%%%%%%%%%%%%%%%%%
In this paper we have mainly studied 
the supersymmetric index of the $SU(N)$ $\mathcal{N}=4$ super Yang-Mills theory on $S^1 \times M_3$.
We have computed the asymptotic behavior of the index in the Cardy limit for arbitrary $N$
by the refined supersymmetric Cardy formula.
We have seen that
the asymptotic behavior of the superconformal index in the large-$N$ limit agrees 
with the Bekenstein-Hawking entropy \eqref{eq:BH} of the rotating electrically charged BPS black hole in $AdS_5$
via the Legendre transformation as recently shown in \cite{Choi:2018hmj}. 
We have also found that the agreement formally persists for finite $N$ 
if we slightly modify the AdS/CFT dictionary \eqref{eq:dictionary} as $\frac{\pi}{2G_N g^3} =4c$.
This implies 
an existence of non-renormalization property for the black hole entropy in the Cardy limit.
We have also studied the cases with other gauge groups and additional matters, and the orbifold $\mathcal{N}=4$ SYM.
It has turned out that
the entropies of all the CFT examples in this paper are given by \eqref{eq:S_final}.

There are several questions and interesting future directions.
Perhaps the most immediate question is whether or not our results match at quantum level.
The first step to test this would be to compute a logarithmic correction to the black hole entropy by one-loop analysis of the supergravity
as in the case of the magnetically charged $AdS_4$ black holes \cite{Liu:2017vbl}.
Our result suggests that the logarithmic correction is absent in the Cardy limit.
It is also of course illuminating to include higher derivative corrections.
Another question is what are physical interpretations of the dominant saddle points  
in the regime ${\rm Re}\left( \frac{i}{\tau\sigma} \right) > 0$, which we have not mainly considered in this paper.
The dominant saddle points in this regime 
are not isolated and technically give the plateau in the function $f(x)$ given in \eqref{eq:fx}
but they are not degenerate at $\mathcal{O}(\beta^{-1})$.
This question might be related to hairy black holes discussed in \cite{Bhattacharyya:2010yg}.
%\cite{Bhattacharyya:2010yg,Dias:2011tj,Markeviciute:2016ivy,Markeviciute:2018yal,Markeviciute:2018cqs}.
It is also interesting to study higher order corrections of $\beta$ to the Cardy limit
in order to interpolate our result to the one in \cite{Benini:2018ywd} which does not take the Cardy limit.
The higher order corrections might be significantly different between large-$N$ and finite $N$.
Another interesting direction is to extend our results for more general holographic 4d CFT 
such as less supersymmetric case.
Perhaps there is an efficient way to compute
the asymptotic behavior of the index especially for class-$S$ theories.

%%%%%%%%%%%%%%%%%%%%%%%%%%%%%%%%%%
%%%%%%%%%%%%%%%%%%%%%%%%%%%%%%%%%%
\subsection*{Acknowledgement}
%%%%%%%%%%%%%%%%%%%%%%%%%%%%%%%%%%
%%%%%%%%%%%%%%%%%%%%%%%%%%%%%%%%%%
The author thanks Nikolay Bobev, Seyed Morteza Hosseini, Seok Kim and Leopoldo A. Pando Zayas 
for comments to the earlier versions of arXiv.
This work has been partially supported by STFC consolidated grant ST/P000681/1.

%%%%%%%%%%%%%%%%%%%%%%%%%%%%%%%%%%
%%%%%%%%%%%%%%%%%%%%%%%%%%%%%%%%%%
%%%%%%%%%%%%%%%%%%%%%%%%%%%%%%%%%%
\appendix
\section{Explicit forms of $V_2 (a)$, $V_1 (a)$ and $\tilde{V}_1 (a)$ for general Lagrangian 4d $\mathcal{N}=1$ theory}
\label{app:general}
%%%%%%%%%%%%%%%%%%%%%%%%%%%%%%%%%%
%%%%%%%%%%%%%%%%%%%%%%%%%%%%%%%%%%
%%%%%%%%%%%%%%%%%%%%%%%%%%%%%%%%%%
Let us consider 4d $\mathcal{N}=1$ SUSY gauge theory with gauge group $G$ 
coupled to chiral multiplets of representation $\mathbf{R}_I$  
having $U(1)_R$ charge $R_I$ and flavor charge $Q_I^j$ of $U(1)_j$ flavor symmetry.
The refined Cardy formula takes the form \cite{DiPietro:2016ond}
\begin{\eq}
I_{S^1 \times M_3}
\underset{\beta \to 0}{\simeq}
e^{-\frac{\pi^2 {\rm Tr}(R) L_{M_3}}{12\beta}} 
\int d^{{\rm rank}G} a\ 
e^{-\frac{\pi^3 i A_{M_3}}{6\beta^2}V_2 (a) +\frac{\pi^2 L_{M_3} }{2\beta} V_1 (a)  -\frac{1}{2\beta} \tilde{V}_1 (a) } ,
\end{\eq}
where\footnote{
More generally, $\tilde{V}_1 (a)$ can have $\ell_{M_3}$ for flavor symmetry background.
For example, $\ell_{M_3}$ for $M_3 =S^1 \times \Sigma_g$ is proportional to magnetic charge
and we have to specify the background magnetic charges.
}
\begin{\eqa}
V_2 (a) 
&=& -\sum_{I\in \rm matters} \sum_{\rho_I \in\mathbf{R}_I}
  \kappa \Bigl( \rho_I \cdot a +\sum_{j\in {\rm flavor}} Q_I^j m_j \Bigr) ,  \NN\\
V_1 (a) 
&=&  \sum_{\alpha\in {\rm root}}\theta (\alpha\cdot a) 
  +\sum_{I\in \rm matters} \sum_{\rho_I \in\mathbf{R}_I} (R_I -1) 
   \theta \Bigl( \rho_I \cdot a +\sum_{j\in {\rm flavor}} Q_I^j m_j \Bigr)  \NN\\
\tilde{V}_1 (a)
&=& \sum_{I\in \rm matters} \sum_{\rho_I \in\mathbf{R}_I}  \rho_I \cdot \ell_{M_3} 
\theta \Bigl( \rho_I \cdot a +\sum_{j\in {\rm flavor}} Q_I^j m_j \Bigr) .
\end{\eqa}

%%%%%%%%%%%%%%%%%%%%%%%
%%%%%%%%%%%%%%%%%%%%%%%
%%%%%%%%%%%%%%%%%%%%%%%
%\begin{thebibliography}{99}
%\bibliographystyle{mybst}
%\bibliography{BH_Cardy}
%%%%%%%%%%%%%%%%%%%%%%%
%%%%%%%%%%%%%%%%%%%%%%%
%%%%%%%%%%%%%%%%%%%%%%%
%\end{thebibliography} 
\providecommand{\href}[2]{#2}\begingroup\raggedright
\endgroup
\end{document}